\documentclass[a4paper]{IEEEtran}
\IEEEoverridecommandlockouts
\usepackage[utf8]{inputenc}
\usepackage{graphicx}
\usepackage{authblk}
\usepackage{hyperref}
\usepackage{amsmath}
\usepackage{amssymb}
\usepackage{amsthm}
\usepackage{graphics}
\usepackage{adjustbox}
\usepackage{float}
\usepackage{relsize}
\usepackage{setspace}
\usepackage{algorithm}
\usepackage[style=ieee]{biblatex}
\usepackage[noend]{algpseudocode}
\usepackage{esvect}
\usepackage{graphicx}
\usepackage{titlesec}
\titlespacing*{\section}{0pt}{1.1\baselineskip}{\baselineskip}

\mathchardef\mhyphen="2D
\makeatletter
\def\subsection{\@startsection {subsection}{2}{\z@}{1ex plus 1ex} {0.5ex plus 0.5ex}{\normalsize\bf\itshape}}
\makeatother

    \title{Generalizing Syndrome Decoding problem to the totally Non-negative Grassmannian }
\author{\IEEEauthorblockN{Kelechi Chuwkunonyerem Emerole\IEEEauthorrefmark{1}, Said Boussakta\IEEEauthorrefmark{1}} \\
	\IEEEauthorblockA{\IEEEauthorrefmark{1}\IEEEauthorrefmark{2} \\
	\IEEEauthorrefmark{1} \IEEEauthorrefmark{2}}}
\date{}
\titlespacing\section{0pt}{12pt plus 4pt minus 2pt}{0pt plus 2pt minus 2pt}
\titlespacing\subsection{0pt}{12pt plus 4pt minus 2pt}{0pt plus 2pt minus 2pt}
\titlespacing\subsubsection{0pt}{12pt plus 4pt minus 2pt}{0pt plus 2pt minus 2pt}
\newtheorem{theorem}{Theorem}
\newtheorem{corollary}{Corollary}
\newtheorem{lemma}{Lemma}
\newtheorem{proposition}{Proposition}
\newtheorem{definition}{Definition}

\newtheorem{keywords}{Keywords}
\newtheorem{remark}{Remark}
\AtBeginDocument{%
  \addtolength\abovedisplayskip{-0.5\baselineskip}%
  \addtolength\belowdisplayskip{-0.5\baselineskip}%
}

\AtBeginBibliography{\small}
\MakeRobust{\vv}
\begin{document}
\setlength{\abovedisplayskip}{0pt}
\setlength{\belowdisplayskip}{0pt}
\setlength{\abovedisplayshortskip}{0pt}
\setlength{\belowdisplayshortskip}{0pt}
\maketitle 
\begin{abstract}
\noindent The syndrome decoding problem has been proposed as a computational hardness assumption for code based cryptosystem that are safe against quantum computing. The problem has been reduced to finding the codeword with the smallest non-zero columns that would satisfy a linear check equation. Variants of Information set decoding algorithms has been developed as cryptanalytic tools to solve the problem.
In this paper, we study and generalize the solution to codes associated with the totally non-negative Grassmannian in the Grassmann metric. This is achieved by reducing it to an instance of finding a subset of the plucker coordinates with the smallest number of columns. Subsequently, the theory of the totally non negative Grassmann is extended to connect the concept of boundary measurement map to Tanner graph like code construction while deriving new analytical bounds on its parameters. The derived bounds shows that the complexity scales up on the size of the plucker coordinates.Finally, experimental results on decoding failure probability and complexity based on row operations are presented and compared to Low Density parity check codes in the Hamming metric.
\end{abstract}

\begin{keywords}syndrome,coding, Grasmannian, complexity , cryptography\end{keywords}

\section{Introduction}
The hardness of decoding the syndrome of a linear code \cite{berlekamp1978inherent} has been useful in designing quantum safe encryption in the Hamming metric using Goppa codes \cite{mceliece1978public} and in the rank metric using Gabidulin \cite{gaborit2014ranksign}. The syndrome decoding problem states that given an instance of parity check matrix $H$, a syndrome of minimum hamming weight $w$ to find a vector $x$ such that $Hx^{T} = s$. The syndrome decoding problem is relevant to the cryptanalysis of code based cryptography. This is because on the input of certain code parameters and with the knowledge of the structure of the code, an attacker can decrypt the ciphertext and reveal the message in the process.Furthermore, this can be done by the Adversary, if it can find the a vector of length $n$ and also if it has the ability to correct $k$ errors. Solutions to the problem in the Hamming metric have been presented using information sets \cite{prange1962use} and its variants \cite{stern1988method} to find the codeword with the smallest weight. Also, these solutions has been extended to  the rank metric to guess the support that contains the error coordinates \cite{gaborit2015complexity}.

The Grassmannian can be divided into positive or negative depending whether the maximal minor of the generator matrix which is the determinant is positive or negative. In other words, a negative Grassmannian has a negative minor while a positive Grassmannian has a positive minor. Furthermore, the positive Grassmannian has positive plucker coordinates as well and the essence of using the positive plucker coordinates as a solution to the syndrome decoding problem is to avoid oscillations that would lead to erroneous results when swapping the columns of the generator matrix. Consequently, in the Grassmann metric, plucker coordinates would replace information sets used in the Hamming metric. \\

However, to the best of our knowledge, no Post quantum based cryptosystem has been designed using codes associated with the Grassmannian in the Grassmann metric. Nevertheless, there is ample evidence that points to the fact there is a connection between the construction of a cryptosystem using a Grassmann based code or a Hamming based code. This is because of the link between the structure of these two codes as explained in this paper \cite{etzion2019grassmannian}. Also, no solution to the problem in the Grassmann metric has been proposed as regards to its use in cryptography. However, for coding applications, research on finding the minimum weight of codewords in the Grassmann metric has been proposed \cite{ryan1990minimum}.

The question of importance moving forward is this, are there codes associated to  Grassmannian varieties with robust theoretical background that can be categorized as a sub family of Tanner graph codes? The synopsis to this question comes from the implication of using Grassmann support and its mathematical framework \cite{gaborit2015complexity} on code based based cryptography in the rank metric. This parameter is actually a parameter used for codes associated to Grassmann varieties. This inspires the paper to connect the dot by expounding on the Grassmann support and its derivatives. Finally, in the theory of toric geometry \cite{postnikov2009matching}, the planar graph that illustrates the totally non negative Grassmannian can be redesigned into a graph similar to a Tanner graph \cite{tanner1981recursive} and possessing the properties of such a graph. Consequently, Non-negative Grassmann codes is a graph based code that can be represented with vertices and nodes just like Tanner graph based codes. 

The solution of the syndrome decoding problem is generalized to the Grassmann metric by using  Plucker coordinate based decoding. This is done by finding the subset of plucker coordinate of codewords of minimum Grassmann weight and with zero error coordinate vectors. This can be seen as a generalization of the birthday attack used in plaintext recovery \cite{stern1988method}. The plucker coordinates of the totally positive Grassmnannian cells are the the columns of the Generator matrix of the code $C(k,n) \subset G_{r}(n,k)$ whose maximal minor is non-zero .
Families of codes associated to Grassmann varieties can be employed in the quantum safe code based cryptosystem because of its efficient decoding procedure \cite{koetter2008coding} and probability to correct low weight codewords \cite{etzion2012codes}.  

The Grassmann graph defines a system of k-dimensional subspaces in an n -dimensional vector space of a finite field of Characteristic 2. The graph also includes a projection of n-k dimensional subspace that form unique pivot positions. These subspaces can be seen as vertices connected by edges, if and only if there is a trivial intersection between the subspaces and in the process producing a unit Grassmann distance. Furthermore, the Graph is characterized by sparse bi-adjacency matrix which can be decomposed into a set of positive Grassmannian Schubert cells \cite{postnikov2009matching}. These cells can be represented by a canonical matrix in a row echelon format with a leading one in each row. The missing element in each row can be modelled using Ferrer's diagram \cite{etzion2012codes} which represents it as partitions. 

The adversary requires knowledge of the map structure in order to decompose the Generator matrix into its row echelon form. In this paper, an instance of a boundary map would be employed to decompose the Generator matrix. They are used to map the $k$ subset elements of the generator matrix into a point in the Grassmannian  in order to find non-negative plucker coordinates with minimum Grassmann distance. Furthermore, an a priori approach can be promoted to find the low Grassmann weight vector by enumerating the basis based on a bound that is expressed as function of the number of positroid cells in the graph $Gr_{k,n}$ with weight $k$.

\subsection{Contribution}\noindent The basic contribution of this paper is to advance the solution of the syndrome decoding problem to the Grassmann metric using Plucker coordinates. First, the theory of plucker coordinates is extended with the transformation of planar graphs to non planar graph with tanner like graph properties. Then, the plucker based decoding based on Gaussian decomposition is presented. Thereafter, analytical bounds on the Grassmann parameters are presented. Finally, Numerical results on the failure probability and the cost of row operations when the solution to the syndrome decoding problem is applied to the Non-negative Grassmann is presented and the result is compared to that of Low Density Parity check codes. 

\section{Preliminaries}
\subsection{Notation}
In this section,a brief summary of some of  the notation used in this paper is provided. $F_{q}$  represents finite field of q elements, $F_{q^{m}}$ represents extension field of degree $m$, $F_{q}^{n}$ represents vector spaces of dimension $n$ over $F_{q}$, $A$  represents $n \times m$ matrix, $a$  represents a  vector,
      $G_{q}(n)$ represents  set of subspaces belonging to $F_{q}^{n}$(Grassmann graph),$E \oplus F$ represents smallest subspace ,$\langle A \rangle$ represents $F_{q}$ span of $A$ 
\subsection{Coding Theory in the Rank Metric}Assuming a bijective mapping between a vector $a$ and a matrix $A \in F_{q}^{m \times n}$, the subspace of a size $n-k$, the complexity of a combinatoric solution is given by $n-k)^{3}m^{3}q^{(n-k)} \biggl [\ \frac{(k+1)m}{n} \biggl ]\ -m$ \cite{gaborit2015complexity}.
Lifting can be performed on an interleaved code by transforming the linear matrix code to a subspace by multiplying its transpose with an identity matrix. The linear matrix code $C [\ m \times n, k ]\ \in F_{q^{m}}$ is a linear code generated by $(m \times n)$ matrices. The linear matrix code can be represented as a function of its basis by $C_{j}=\sum_{i=1}^{m}X_{ij}\beta_{i} \forall j \in \{1, \hdots, n \}$ where $\beta_{i}$ is a basis of a subspace $F$ over $F_{q^{m}}$. The basis of a subspace over $F_{q}$ multiplies $C$ by a non zero element which does not affect the rank distance between codewords. The basis can also be a row of a generator matrix $G \in F_{q^{m}}^{k \times n}$ which has the complexity of $k(n-k)m^{2}log_{2}qbits$ \cite{chabaud1996cryptographic}. The dimension of the subspace determines the weight of the codeword and the number of subspaces is given by the Gaussian coefficient expressed as \\
\begin{equation}
    \begin{pmatrix}n\\w \end{pmatrix}_{q}= \prod_{i=0}^{k-1}\frac{q^{n}-q^{i}}{q^{w}-q^{i}}
\end{equation}
$w$ is the weight and $q^{m}$ and $q^{i}$ are monomials over $F_{q^{m}}$.\\
In information set decoding, the probability of finding the codeword given a $[\ n,k,t+1 ]\ $matrix code is given by \\
\begin{gather}
    P_{dec}=\frac{\biggl( \begin{matrix}n-k\\t \end{matrix} \biggl)}{\biggl( \begin{matrix}n\\t \end{matrix} \biggl)} 
\end{gather}
with complexity $P_{dec}=O(1).2^{nH_{2}(t/n)-(1-k)H_{2}(t/(n-k))}$
where $H_{2}(x)=-xlog_{2}(x)-(1-x)log_{2}(1-x)$ \cite{kachigar2017quantum}.If the parity check matrix $H$ is expressed with respect to $(n-k) \times n$ identity matrix, an $m \times k$ zero matrix and $(n-k-n) \times k$ random matrix code chosen uniformly as $H=(I/0/R)$ then the linear matrix code is called a simple code and to decode such a matrix value when $m < \frac{m+n-\sqrt{(m-n)^{2}+4km}}{2}$ is given by $P_{f} \sim \frac{1}{q^{m-w+1}}$ as $q \rightarrow \infty$.\\
The bound on the weight of the error vector is given by the Gilbert-Varshanov bound \cite{varshamov1957evaluation} which is defined as thus\\
\begin{definition}
The number of elements of a sphere $S$ given integers $n,m,q,t$ with radius $t \in F_{q^{m}}^{n}$ is equal to the number of spaces with $m \times n$ bases of dimension $t$. For $t \geq 1$ this follows that \\
\begin{equation}
 S= \prod_{j=0}^{t-1}\frac{(q^{n}-q^{j})(q^{m}-q^{j})}   {q^{t}-q^{j}}   
\end{equation}
\end{definition}
For a ball of radius $t$, the volume of $B= \sum_{i=0}^{t}S(i)$.Also for a matrix code $C$, if $B \geq q^{m(n-k)}$ and $\mathlarger{\mathlarger{{\sum}}_{j=0}^{d-2}} \biggl ( \begin{matrix}n-1\\j \end{matrix} \biggl  ) < 2^{n-k}$ then the smallest integer $t$ is referred to as the Gilbert-Varshanov bound.

\subsection{Syndrome Decoding Problem}
The Syndrome decoding problem is defined here in terms of complexity theory
\begin{definition}
The a priori probability of finding a codeword $x_{i}$ with non-zero codewords $ \leq w$ and an integer which represent the $ith$ column of an error  applied to a Code $C$ which transforms it to $C^{'}$, and in the process satisfying the expression $H^{T}x=s$, where $s \in_{R} F_{q^{m}}^{n-k}$ is a syndrome and $H$ is a parity check matrix over $F_{q^{m}}$.\end{definition} 

Consequently, to generalize this problem to the Grassmannian metric, it has to be reduced to an instance of finding the plucker coordinates of codewords with lowest Grasmann weight.
\begin{definition}
Let plucker coordinates be denoted as  $\Delta_{I,J}(G) > 0$ which forms the columns of the generator matrix. The syndrome decoding problem is to find linearly dependent subset of plucker coordinate with $w$ columns such that $G_{i,j-k} \wedge v_{j} = u_{i}$ were a basis $B$ is defined thus; $B= \{ u_{i},v_{j} | i \in I,j \in J \}$, a $k \times n-k$ matrix $M_{v}$ and a $k \times n-k$ generator matrix $G$ with rank $K$.
\end{definition}

\subsection{Grasmmaninan theory} 
\begin{definition}
Totally non-negative Grassmanninan \cite{postnikov2009matching} is the point in the Grassmann graph with positive plucker coordinates $\Delta_{I} \neq 0$
\end{definition}
In other words its maximal minor is positive and it can combinatorially analyzed using planar bipartite graph. The matroid of the totally positive Grassmannian is termed a positroid.
\begin{definition}
The boundary measurement map \cite{postnikov2009matching} is defined as $b:R_{>0} \rightarrow G_{L_{k}}.A \in Gr_{n,k}$ where $A$ is a $k \times n$ biadjacency matrix with a rank $k$ which are represented by incoming boundary edges and the map depends on the coloring of the vertices.
\end{definition}
The matrix has a maximal minor $\Delta_{I}=1$ that forms the plucker coordinates on $Gr_{n,k}$ with column vectors $\frac{I}{A}$ that gives the basis of the subspace. 
Furthermore, the coordinates of $A$ can be defined as follows with slight abuse of notation
$\varphi(A) = \langle ( u_{i} + \sum_{j=1}^{n-k} A_{i j} v_{j} ) \rangle \forall 1 \leq i \leq k$.

$R_{>0}$ is characterized by the set of all the biadjancency matrix A. The subspace in this set is a graph of a map from a projection to its orthornormal that is $V \rightarrow V^{\perp}$ and direct sum expression given by $V \oplus V^{\perp} \cong R^{n}$ with a basis $V= \{v_{1}, \hdots, v_{a} \}$. 

Let the map of a subspace $U$ to its local diffeomorphism be given as $\phi(u)=(\phi_{1}, \hdots,\phi_{n})(u_{1},\hdots,u_{k})$, then it follows that the tangential space at any point of the map has a basis with coordinates $\{ \frac{\partial \phi}{\partial u_{1}}, \hdots, \frac{\partial \phi}{\partial u_{k}} \}$. In other words, the tangential space can also be represented by the derivative of the Grassmann.If there is an open subspace in the Grassmann graph $Gr_{n,k}$, then we have $U= \{W : W \cap V^{\perp}= \{0\} \} \subset R^{k} \times R^{n-k}$ for any $W \in U$.

There are complex numbers $c_{ij}$ such that $v_{i}+\sum_{j=1}^{b}c_{ij}v_{j} \in W$ which is linearly isomorphic. Therefore, the graph becomes $U(S)= \{ v+ Sv:v \in V \}$ such that $v \mapsto (V,S(v))$. If $v=0$, then $U(C)=0$ from the nullity of maps. If $V$ is decomposed to subspaces $P$ and $Q$ where $Q \in U_{A}$ and $U_{A}$ is a set of all subspace $P \subset V$ such that $V \cap U_{A}= \{0 \}$, then we have $P=(P \cap Q) \oplus P^{'})$ for some $P^{'}$ isomorphic to $P/(P \cap Q)$. 

Furthermore, for a direct sum decomposition, the intersection of $P$ and $Q$ is trivial which now becomes $P + ((P \cap Q)\oplus P^{'})=P \oplus Q^{'}$. If the subspace $E$ is decomposed, we now have $E=(E \cap V) \oplus E^{'}$ for some $E^{'} \subset R^{n}$ where the intersection $E \cap V$ tends towards the solution \cite{koetter2008coding}. 

Finally, an injective transformation $F_{k}(V)$ given by $T:R^{k} \mapsto V$ is an open subset of $L(R^{k},V)$ and a space with $dim(F_{k}(V)=kn$. In other words, $F_{k}(V)$ is the projective geometry of $V$ and its quotient space generates the Grassmannian space.

\begin{proposition}
Let $V$ be a linear subspace and $V^{\perp}$ its orthonormal projection. Let $U_{A}$ be a set of all projections $P_{V} \subset V$ through a map $U = v + Sv$. Then $U_{A}$ lies in $L(V,E)$, if a linear isomorphism $T \in \pi^{-1}(U_{A})$ exists. \end{proposition}.
\begin{proof}
If there is an open subspace in the Grassmann graph $G_{n-k}$, then $U= \{E \cap V^{\perp} = \{ 0 \} \}$ and $U(S)=\{v + Sv:v \in V \} : v \mapsto (v,S(v))$ where a subspace $S \subset V \oplus E$. This implies that $S \cap E = \{0 \}$. Lets define two  projections $P_{V^{'}} : V^{'} \mapsto V$ and $P_{V} : V \mapsto V^{'}$ where $P_{V}(v)$ is related to $P_{V^{'}}$ by the expression $P_{V}(v) = (P_{V^{'}})^{-1}(v)-v$. Given $U_{A}$ a set of all projections $P_{V} \subset V$, we have a linear isomorphism $T \in \pi^{-1}(U_{A})$ and a projective geometry $F_{K}(v)= \pi^{-1}(U_{A})$ where $\pi^{-1}$ is an invertible function. Then it follows that the intersection of $T$ and the biadjacency $A$ is trivial that is $\pi(U_{A} \cap A = \{0 \}$, if the function $\pi$ can be inverted and if a map $f(T)=0$. For $v \in V$, it is assumed that the k dimensional subspace is equivalent to its transformation for some $v^{'} \in V$ that is $v + S(v) = v^{'} + S^{'}(v^{'})$. It follows that $v-v^{'}=S^{'}(v^{'})=S(v) \in E \cap V^{\perp}= \{0 \}$,  $\implies S(v)=S^{'}(v^{'})$. Concatenating the linear isomorphism $T$ with the projections $P_{V^{'}}$ and $P_{V}$, we have $f_{T}(v)=(P_{V^{'}} \circ T) \circ (P_{V} \circ T)^{-1} \forall v \in V$ and if $f$ restricts $S = S^{'}$ on $L(V,E)$ then it becomes $f_{T} : \pi^{-1}(U_{A}) \mapsto L(V,E)$ $\implies$ that $P_{V}(v)=(P_{V^{'}})^{-1}(v)-v=v+Sv$. This results to $P_{V}(v)=Sv$ and $f_{T}(v)=(P_{V^{'}} \circ T) \circ (P_{V} \circ T)^{-1} = id_{V,V^{\perp}}$
\end{proof}
\section{Extending the theory on Non negative Grassmann}In this section, we would try to link the totally non negative Grassmann to tanner code like constructions by transforming it from its planar structure to non planar structure. This can be seen as intersecting the theory of distance transitive graph and coding theory based on the framework of Grassmann variety. First, we redefine the concept  of boundary measurement maps and thereafter present a logical breakdown of how this map can be represented as a binary matrix. The boundary measurement maps are designed as a mapping or transformation of vertex set in a planar bipartite graph to edge weights defined as a set of vertices in a cell in the Grassmannian graph. Given a set $I_{f} \subset I$, removing an element from the set, an embedding can be constructed from the bipartite to the Grassmannian as $G_{r_{k,n}}(R) \rightarrow RP^{\biggl (\begin{matrix}n \\ k \end{matrix}\biggl )-1}$ which forms a guage transformations expressed  as a function of matroids $Meas:R_{>0} \rightarrow G_{r_{k,n}}(R)$ where $G_{r_{k,n}}(R)$ is  $k$ planes on an n-dimensional space which is not affected by the ratios of $k \times k$ minors of a $k \times n$  code. To decompose the Grassmann, an arbitary edge function is selected such that $e:u \rightarrow v$ and if the vertex is coloured, another edge function is selected $e^{'}: v \rightarrow w$ by maximum revolution.Depending on the coloring, this maximum revolution can be clockwise or anticlockwise. This maximum revolution induces self intersections through the path and can define the boundary measurement as $M_{ij}=\sum_{P:e \rightarrow e^{'}}(-1)^{wind(R)}wt(P,y)$ where the factor $(-1)^{wind(R)}$ is bound by the number of connection between sources to the planar bipartitte graph which is made up of $n$ external nodes of perfect orientation and $k$ sources of perfect orientation and $wt(P,y)$ is the weight of the path.

The planar bipartite graph structure with perfect orientation\cite{franco2014bipartite},\cite{postnikov2009matching} would be employed to buttress the idea. This is shown in Figure 1 and Figure 2. First, the planar bipartite graph is transformed into non planar bipartite graph taking note of the sources and external nodes while labelling them accordingly for convenience purposes. If the row and column are of the same node, the code entry is set to 1,if there is no path connecting the nodes, the map code entry is set to 0. Finally, the condition in literature is modified to support the objective of the idea by stating that if there is a negative sign then the entry is set to 0 and set to 1 if otherwise. Therefore,a boundary measurement mapping $A$ and $B$ produces the Grassmannian $G_{r>0}(2,4)$ and $G_{r>0}(2,6)$ respectively which is constructed using the flows as regards to whether it is clockwise or anticlockwise as follows; \\
\begin{figure}[!htb]
    \centering
    \includegraphics[width=9cm, height=7cm]{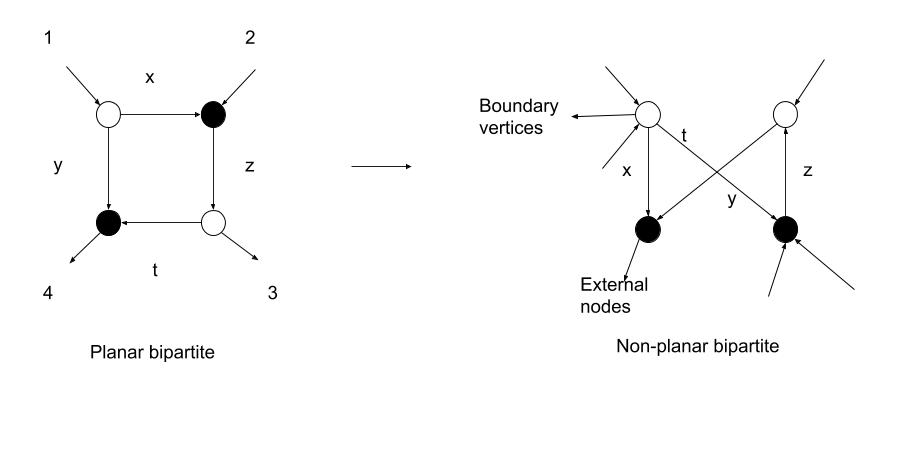}
    \caption{Non planar bipartite graph with perfect orientation containing 2 boundary vertices, 2 external nodes and a face transformed to its non planar structure}
    \label{fig:my_label}
\end{figure}
\begin{figure}[!htb]
    \centering
    \includegraphics[width=9cm, height=7cm]{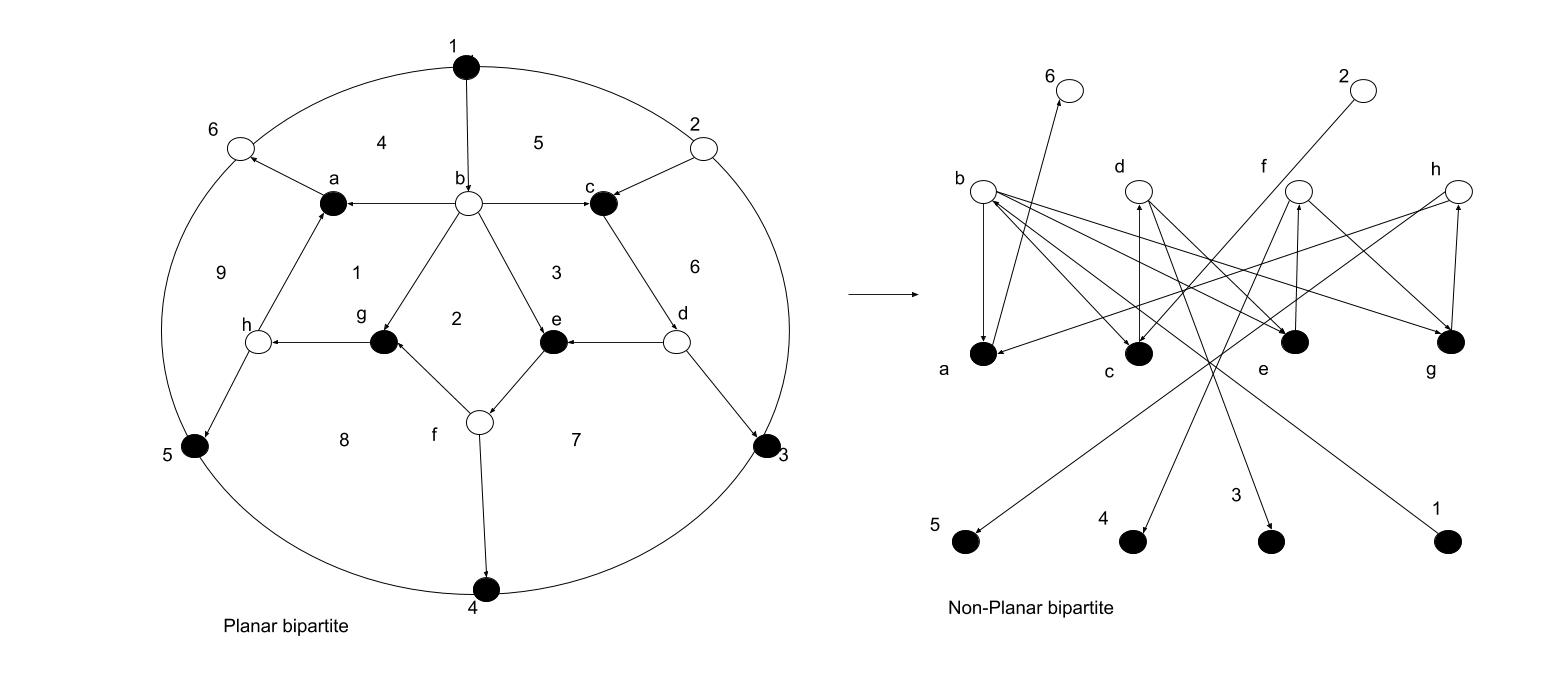}
    \caption{Non planar bipartite graph with perfect orientation containing 2 boundary vertices, 6 external nodes and 9 faces transformed to its non planar structure }
    \label{fig:my_label}
\end{figure} \\
\begin{gather}
A = \begin{bmatrix}1 & 0 & -t+x & -(y+xzt) \\              
                       0 & 1 &   y  &    zt    \end{bmatrix} \implies \begin{bmatrix}1 & 0 & 0 & 0 \\ \nonumber
                                                                              0 & 1 & 1 & 1  \end{bmatrix} \\ \nonumber \rightarrow G_{r>0}(2,4) \\
\end{gather}
The same procedure is extended to $B$ as well \\
\begin{gather}
B= \begin{bmatrix}1 & 1 & 0 & 0 & 0 & 0 \\ \nonumber
                1 & 0 & 0 & 0 & 1 & 1  \end{bmatrix}  \rightarrow G_{r>0}(2,6) \\   
    \end{gather}
The dimension of the Grassmanian parametrized from $G_{r>0}(2,4)$ is given as $4$, then the number of boundary vertices $k$ is computed as follows $k(n-k)=4;k=2$ while that of the Grassmannian parametrized $G_{r>0}(2,4)$ is given as $6$, then the number of boundary vertices $k$ is computed as follows $k(n-k)=6;k=2$

For a set $I=\{1,2 \}$ and a minor $J={2,6}$,a modified  plucker coordinate for $\Delta_{2,6}$ can be computed as follows \\
\begin{gather}
    \Delta_{26}=f/g=\frac{(1b+C2)(1b+ab)}{1+C2}
\end{gather}

\section{Decoding with Plucker coordinates}In this section we present the idea of decoding with plucker coordinates as a solution to the Syndrome decoding problem in the Grassmann metric. It is pertinent to note that this method is analogous to an optimized variant of Information set decoding.\\ Let $C \subset G_{r}^{+}(n,k) \in F_{2}^{k + l}$ be a code associated to the totally non-negative Grassmannian with a generator matrix $G \in F_{2}^{(k+l) \times l}$ and a subset of the matroid space $Mat$., we have $G = \begin{pmatrix}g_{0}            & g_{1}           & g_{2}           & \dots & g_{n}    \\
                                                 g_{0}^{q}        & g_{1}^{q}       & g_{2}^{q}       & \dots & g_{n}^{q} \\
                                                 \vdots           & \vdots          & \vdots          & \dots & \vdots    \\
                                                 g_{0}^{q^{k-1}}  & g_{1}^{q^{k-1}} & g_{2}^{q^{k-1}} & \dots & g_{n}^{q^{k-1}}
\end{pmatrix}$. 
The element of the Grassmannian are the linear span of the columns of the generator matrix which produces the subspace $V = \langle g_{i}, \hdots,g_{n}^{q^{k-1}} \in R^{k} \rangle $ and the linear span of the rows of the generator matrix produces the subspace $U = \langle g_{i}, \hdots, g_{n} \rangle \subset R^{n} $.
By employing Gaussian elimination and taking an instance of the boundary map $\tau \in b$, we generate an equivalent code $C^{'} = \tau(C)$ with generator matrix $G^{'}$ in row echelon form $G^{'} =                         \begin{pmatrix} 
I^{l} & O^{l} & H^{'} \\
O^{n-k-l} & I^{n-k-l} & H^{''} 
\end{pmatrix}$ were $H^{'} \in F_{2}^{(k+l) \times (k+l)}$,$H^{''} \in F_{2}^{(2k+l) \times (k+l)}$ and $I^{n-k-l}, I^{l}$ are  identity matrices of size $n-k-l$ and $l$ respectively. $0^{n-k-l}, 0^{l}$ are zero matrices of size $n-k-l$ and $l$ respectively. Select plucker coordinates $\Delta_{I,J}(G)$ with size $k+l$ for $H^{'}$ and another plucker coordinate $\Delta_{I,J}$ for $H^{''}$ were $I = \{i_{1} <, \hdots, < i_{k} \}$ are $k$ elements of $G$. Applying cycle shift to the columns of $H^{'}$ and removing indices $i \in I$ to form a basis of the subspace $V^{'}= \langle g_{2}, \hdots (-1)^{k-1} g_{n}^{q^{k-1}}, g_{1}$ and also cycling shifting the columns of $H^{'}$ and removing indices $i \in I$ to form the basis of the extended subspace $U^{'}=\langle g_{2}, \hdots, g_{n},g_{i} \rangle$. A linear combination of the the $k-1$ columns of the subspace $V^{'}$ will form a vector $\tau(V^{'})$ and a linear combination of the $n$ columns of the subspace $U^{'}$ will form a vector $\tau(\Delta_{U^{'}})$ with a pivot centered around $\tau \in b$. Add $\tau(V^{'}) + \tau(\Delta_{U^{'}})$ and check if the Grassmann weight $d(V^{'} \cap U^{'}) \leq w - n + k-1$ and stop. if the last condition is not met, then the process is repeated. It can be said that if the cyclic shift is applied, $I$ becomes $I^{'}$. The Gaussian decomposition operation is a function of the ordering of the plucker coordinate vectors.  \\
\subsection{Correctness}
 The identity matrix $I_{u}$ and the zero matrix $O_{V}$ were both are restricted to $n-k-l$ plucker coordinate positions, $I_{U}=\begin{bmatrix}I^{l} \\I^{n-k-l}\end{bmatrix}$ and $O_{U}=\begin{bmatrix}O^{l} \\O^{n-k-l}\end{bmatrix}$. We transform the matrix $I_{U}$ and $O_{V}$ by multiplying by the parity check matrix $H$ as follows
$I_{U}H=\begin{bmatrix}H^{'} & I^{l} \\H^{''} & I^{n-k-l}\end{bmatrix}$ and $O_{V}H=\begin{bmatrix}H^{'} & O^{l} \\H^{''} & O^{n-k-l}\end{bmatrix}$. Furthermore, multiplying the error vector $x$ to both matrices were $x$ is generated by $k+l$ entries $I_{U}Hx^{T}=\begin{bmatrix}H^{'}x^{'T} +x^{''T} \\H^{''}x^{'T}+x^{''T}\end{bmatrix}$ and $O_{V}Hx^{T}=\begin{bmatrix}H^{'}x^{'T} \\ H^{''}x^{'T}\end{bmatrix}$. Concatenating the matrices becomes \\
\begin{gather}
I_{U}O_{V}Hx^{T}=\begin{bmatrix}(H^{'}x^{'T}.H^{'}x^{'T})+(H^{'}x^{'T}.x^{''T})\\(H^{''}x^{'T}.H^{'}x^{'T})+(H^{''}x^{'T}.x^{''T})\end{bmatrix}
\end{gather}
let $s=(s^{'},s^{''})$ be the coordinate of the syndrome then \\
\begin{gather*}
    I_{U}O_{V}s^{T}=\begin{bmatrix}H^{'}x^{'T}+O^{l}\\H^{''}x^{'T}s^{'}+O^{l}\end{bmatrix}= \begin{bmatrix}H^{'}x^{'T}\\H^{''}x^{'T}s^{'}\end{bmatrix}
\end{gather*}

\noindent Let $B(k,n)$ be the plucker coordinate of all subspaces with restriction in the first $k$ plucker coordinates $g_{1}, \hdots g_{k}^{q^{2k-n}}$. The $k \times k$ minor $\Delta_{B(n,k)}$ of the generator matrix $G^{'}$ is the set of $k$ plucker coordinates in $G_{r}^{+}(k,n)$. The instance of the boundary measurement map is validated by the Adversary on the condition that $ \Delta_{B(n,k)}(G) \neq 0$. It can be said that $B(k,n)$ which is the bounded affine permutations constitute the set of information sequences. The instance of the boundary measurement map can be represented by a Vandermonde matrix such that the plucker coordinate is the column set of $I^{n-k-l} \in G^{'}$. Afterwards, the adversary selects an arbitrary subspace $V$ with basis $V = \langle 0, v_{1}, \hdots v_{k+t} \rangle \subset C^{'}$ and choose the codewords with minimum weight $w \leq q^{\frac{k(k-1)}{2}}$. Finally, the Adversary checks if $d(U \cap V) \leq w $ and stops. By induction, it can be seen that there are $q^{\frac{k(k-1)}{2}}.\biggl [ \begin{matrix} k \\ r \end{matrix} \biggl ]_{q}$ ways of choosing the basis of the subspace $V$ and $q^{\frac{k(k-1)}{2}}.\biggl [ \begin{matrix} n-r \\ k-r \end{matrix} \biggl ]_{q}$ ways of choosing subspace $U$.The proof of this claim is presented in Theorem 3. Therefore the probability of guessing correctly the error free plucker coordinates is given as $\frac{\biggl [ \begin{matrix} n-r \\ k-r \end{matrix} \biggl ]_{q}}{\biggl [ \begin{matrix} k \\ r \end{matrix} \biggl ]_{q}}$.

\section{Analytical Bounds on Grassmann parameters}

\begin{proposition}
Let $U,V \in F_{q^{m}}$. As $q \mapsto 1$ and defining a map $P_{v}:F_{q}^{n} \mapsto F_{q}^{n-1}/V^{'}$ then $d(U,V) \leq 2q \bigl [\  \begin{matrix}n \\ k \end{matrix} \bigl ]\ _{q} $
\end{proposition}
\begin{proof}
$k$ subspaces $U,V$ of $F_{q^{m}}, d(U,V)=k-dim(U \cap V)$ and for vector spaces over the same field, we have $dim(V \cap G)=dim(V)+dim(G)-dim(V.G)$, therefore it follows that \\ $d(U,V)=k-(dim(U)+dim(V)-dim(U.V) \leq k-(k+k-(k-r)=r$. Given a subspace with dimension $k$, $\biggl [\ \begin{matrix}n \\ k \end{matrix} \biggl ]\ _{q} = \prod_{i=0}^{k-1}\frac{q^{n}-q^{i}}{q^{k}-q^{i}}$, Selecting a $k-1$ dimensional subspace $V^{'}$ of $F_{q}^{n-1}$ to construct an arbitrary $k$ dimensional subspace such that $V \cap V^{'} = \{ 0 \}$. Selecting a basis $v \in V^{'}, v^{'} = \{ v_{1} < \hdots v_{k-1} \} \subset N$ of a linear map defined thus $P_{v} : F_{q}^{n} \rightarrow F_{q}^{n-1}/V^{'}$ to construct a bundle $\phi^{-1}(1) = V$. If $dim V^{'} = r$, then the number of bundles is equivalent to the number of enumerated bases of size $\{1, \hdots, n-k \}$ over $F_{q}$ which is $q^{n-k}$. This results to the identity \\

\begin{gather}
\biggl [ \begin{matrix}n \\ k \end{matrix} \biggl ]_{q} = \prod_{i=0}^{k-1}\frac{q^{n}-q^{i}}{q^{k}-q^{i}} = \biggl [ \begin{matrix}n - 1 \\ k \end{matrix} \biggl ]_{q} + q^{n-k} \biggl [ \begin{matrix}n - 1 \\ k - 1 \end{matrix} \biggl ]_{q} \\
\end{gather}
this follows that for $0 < k < n$ \\
\begin{gather}
  \leq \frac{q^{n-1}-1}{q^{k}-1} + q^{n-k}. \frac{q^{n-1}-1}{q^{k-1}-1} \\
  \leq \frac{q^{n-1}-1}{q^{k}-1} +  \frac{(q^{n-k})(q^{n-1}-1)}{q^{k-1}-1}
 \end{gather}
Using a generalized identity \cite{andrews1986q} and doubling the right hand side of Equation (13), vectors except one of the $q$ multiples of $v$ can be computed as
\begin{gather}
 \biggl [ \begin{matrix}n \\ k \end{matrix} \biggl ]_{q} = \sum_{i=0}^{k-1}q^{(n-k)(k-i)} \biggl [ \begin{matrix}n-i \\ i \end{matrix} \biggl ]_{q} \leq
\end{gather}
\begin{gather}
    \prod_{i=0}^{k-1}\frac{q^{n-i+1}-q}{q^{i}-1} = \\ \nonumber
    \end{gather}
    factorize $q$ based on cardinality \cite{gabidulin2015subspace} it becomes
    \begin{gather}
    \prod_{i=0}^{k-1}q\frac{q^{n-i}-1}{q^{i}-1}  \\ \nonumber
    \end{gather}
\end{proof}
\begin{remark}
Proposition 2 gives the bound on the total number of error patterns with $k$ errors that the enumerator can compute given the size of the plucker coordinate as $\biggl [ \begin{matrix}n \\ k \end{matrix} \biggl ]_{q}$.
\end{remark}
\begin{lemma}
The basis of the concatenation of the subspace $U$ and $V$ induces a subgraph with no cycle whose weight of its total path is equivalent to the plucker coordinate of the Grassmannian graph.
\end{lemma}
\begin{proof}
Given a bounded permutation $f_{x}(i)=min \{y \geq i/v_{i} \in span \{v_{i+1},v_{i+2}, \hdots v_{j} \} \}$ where $v_{i}$ are the columns of the arbitrary  space of $S$, taking basis $\{v_{i+1},v_{i+2},\hdots v_{j} \} \}$ and extend it to $U \cap V$ as follows $v_{i+1},v_{i+2}, \hdots,v_{j},e_{i-m+1},\hdots,e_{i}$ and $\{v_{i+1},\hdots,v_{j},f_{i-m+1},\hdots,f_{k} \}$ through the path of the disk divided by a face $f \in U$ then we have $P = \{e_{i-m+1},f_{i-m+1},\hdots,e_{i},f_{i} \}$ which forms a basis. The plucker coordinate now becomes $\Delta_{I}(G)=\sum\prod_{P_{i}} wt(P_{i})$, which implies that$\Delta_{I}(G)$ divides the vertex set $\Delta_{I}$ indexed by $I$ an identity matrix such that each elements $e \in E$ and $f \in F$ induces a subgraph in $\Delta_{I}$  
\end{proof}
\begin{theorem}
the intersection array is given by $b_{r_{k}} \leq q^{i(i-1)/2}[\ \begin{matrix}n \\ k \end{matrix}]\ _{q} $.
\end{theorem}
\begin{proof}
Connecting $k$ to $k+1$ vertices with a rank $r$ will give the boundary measurement map transformation from the planar bipartite graph $G$ to non-planar Grassmannian $G_{r}$ if $k+1 \in I$ 
For $k \not \in I$ and with plucker coordinates given as $\Delta_{I}(G)=\Delta_{I}(G_{r})+r\Delta_{I}- \{ k+1 \} \cup \{ k \}(G_{r})$ this implies that $\Delta_{(I \ \{ r \}) \cup \{ k \}}=(-1)^{t}b_{r_{k}}\geq 0 $ where $t=|I \cap [\ r+1,k-1]\ |$ resulting in the probability
\begin{gather}
    (-1)^{i}\prod_{j=1}^{i}\frac{q^{j-1}q{m-i+1}-1}{q^{i}-1}=(-1)^{i}q^{i(i-1)/2} [\ \begin{matrix}m \\ i \end{matrix}]\
\end{gather}
\end{proof}
\begin{remark}
 It can be seen from Theorem 1, that the intersection array depends on the degree of the extension field $m$. Increasing the degree extension of the field or the power of the prime increases the intersection array of the Grassmannian graph. Also, each row operation of the Gaussian elimination process preserves the intersection array of the graph. Furthermore, this increases the size of the plukcer coordinates thereby reducing the complexity of the solution.
\end{remark}
\noindent Assuming two codewords $C_{1}$ and $C_{2}$ have rank weight $k_{1}$ and $k_{2}$. $C_{1}$ and $C_{2}$ have two different subspaces $V$ and $U$ where $V= \{ v_{1}, \hdots, v_{k_{1}} \}$ and $U= \{u_{1}, \hdots, u_{k_{2}} \}$, then the product of the spaces is bounded by $\langle VU \rangle \leq k_{1}k_{2}$ where $k_{1}$ and $k_{2}$ are the dimensions of the spaces $V$ and $U$. if $k_{1}k_{2} < m$ then the probability holds $Pr(dim\langle VU \rangle < k_{1}k_{2} \leq \frac{q^{k_{1}k_{2}}}{q^{m}}$. This probability is the probability of enumerating the bases in order to find the candidate codewords given the dimension\\

\begin{corollary}
If $A$ is random and $B$ is fixed then the probability that a space $U$ and $A$ a base that generates a random space with dimension $k_{1}$ is at least $1-k_{1}\frac{q^{k_{1}k_{2}}}{q^{m}}$ where $\text{dim}\langle AU \rangle = k_{1}k_{2}$. 
\end{corollary}
\begin{proof}
There exist a codeword $C \in U$ where $U$ is a space and $C \notin F_{q}$, then given $\text{dim}\langle AU^{2} \rangle=k_{1}k_{2} $ and an error $e \in \langle AB \rangle$ with $e \notin A$ then the product $CU$ is an element of the space $U$.
\end{proof}
\begin{theorem}
Let $A$ be a base that generates a fixed space with dimension $k_{1}$ and $B$ a base that generates a random space with a basis such that dimension $k_{2}^{'}=k_{1}^{'}(1-k_{2})$, if $A \cup \langle AB \rangle =\beta$ with its probability of enumeration given as $1-k_{2}\frac{q^{2k_{1}k_{2}^{2}+k_{2}(k_{2}+1)}}{q^{m}}$ holds.\\
\end{theorem}
\begin{proof}
We have $\cap_{i}\beta_{i}^{-1}s=A$ then  $A\cup \langle AB \rangle= \beta$ where $\langle AB \rangle$ is the product of the space with their attendant bases $A$ and $B$ which gives a new basis $\beta$. If $A$ is random the dimension becomes $k_{1}^{'}k_{2}-k_{2}=k_{2}(k_{1}^{'}-1)$ then a random space with a base $B$ has a dimension $k_{2}^{'}=k_{1}^{'}(1-k_{2})$ as given. If $\langle AB \rangle \cap \langle AB \rangle_{-1}=A$ such that the dimension of a fixed base $B$ is      $    \text{dim}B=\text{dim}(k_{2})+B\beta^{-1}$ which is equivalent to $\frac{k_{2}(k_{2}+1}{2}+B\beta^{-1}$. Multiplying both sides by $2$ now becomes $k_{2}(k_{2}+1)+2k_{1}k_{2}^{2}$ with the given probability
\end{proof}
\begin{remark}
 Corollary 1 shows the probability of finding the codewords in plucker coordinates embedded in a space of dimension $k_{1}$ when the Code associated to the totally Nonnegative Grassmann is concatenated with a subspace generated by a random basis.\\ Theorem 2 takes it further by describing the probablity of finding the codeword if the subspace is a linear span of a fixed basis and a random basis with random coordinate vectors. It can be seen that the probability scales with increase in the $k_{2}$ positions the decoding algorithm searches for.
\end{remark}
\begin{theorem}
Given $U,V \in G_{r}(n,k)$ and $d(U,V)=\texttt{dim}(U) + \texttt{dim}(V)-2 \texttt{dim}(U \cap V)=k-r$ where $k$ is the dimension of the subspace and $r$ is the rank with integers $l,p,m$ then the bound from the Gaussian coefficient on $d(U,V)$  given by
\begin{equation}
    \biggl [ \begin{matrix} n \\ k \end{matrix} \biggl ]_{q} = \sum_{k=0}^{\infty}\frac{q^{\frac{k(k-1)}{2}}}{(1-q)(1-q)^{2}\hdots (1-q^{k})}.\biggl [ \begin{matrix} n-r \\ k-r \end{matrix} \biggl ]_{q}.\biggl [ \begin{matrix} r \\ k-m \end{matrix} \biggl ]_{q}.\biggl [ \begin{matrix} k \\ r \end{matrix} \biggl ]_{q}
\end{equation}
\end{theorem}
\begin{proof}
 Starting with a basis for $U$, $B_{1}=(e_{1}, \hdots, e_{m}) $, picking randomly linearly independent vector $x_{U_{i}} \in U$. Then search for a coordinate of $x_{U_{i}}$ and replace to produce a new basis for $U$ after repeated procedures to give $B_{1} = e_{1}^{'}, \hdots e_{m}^{'},x_{U_{1}}, \hdots x_{U_{k}}$ and update count as
\begin{gather}
\texttt{Count}_{U}=\prod_{k=0}^{U_{i}-1}q^{k}=\sum_{k=0}^{U_{i}}q^{\frac{k-1}{2}}\biggl ( \begin{matrix}n \\ k \end{matrix} \biggl )_{q}. 
\end{gather}
Then the same process follows for $V$  a basis, $B_{2}=(f_{1}, \hdots, f_{m})$ is selected. Then, random linearly independent vectors $y_{V_{i}} \in V$ is selected as well and a search for coordinate of $y_{V_{i}}$ is conducted which is now replaced to produce a new basis for $V$ after repeated procedures to give $B_{2}=f_{1}^{'}, \hdots f_{m}^{'},y_{V_{1}},\hdots y_{V_{k}}$ and updating the count gives
\begin{gather}
\texttt{Count}_{V}=\prod_{k=0}^{V_{i}-1}q^{k}-q^{k-r}=\sum_{k=0}^{V_{i}}q^{\frac{k(k-r)}{2}}\biggl ( \begin{matrix}k \\ r \end{matrix} \biggl )_{q}
\end{gather}
Then, finally starting with a basis for $U \cap V$, $B_{3}=(g_{1},\hdots, g_{m})$, then another random linearly independent vector $z_{i} \in U \cap V$ is selected to produce a new basis after repeated procedures $B_{3^{'}}=(g_{1}^{'},\hdots g_{m}^{'},x_{U_{1}}, \hdots , x_{U_{k}}$ and $B_{3^{''}}=(g_{1}^{'},\hdots g_{m}^{'},y_{V_{1}}, \hdots , y_{V_{k}})$. Sampling an integer $l_{i} \in L$ where $L=\texttt{Vect}(x_{U})$ and $p_{i} \in P$ where $P = \texttt{Vect}(y_{V})$ and updating the count as 
\begin{gather}
\texttt{Count}_{*}=\prod_{k=0}^{U_{i}-V_{i}-1}q^{k}-q^{k-r+t}-q^{k-r+p}= \\ \nonumber \sum_{k=0}^{U_{i}-V_{i}-1}q^{\frac{k(k-r)}{2}}\biggl [ \begin{matrix} n-r \\ k-r \end{matrix} \biggl ]_{q}. \\ \nonumber \biggl [ \begin{matrix} r \\ k-t \end{matrix} \biggl ]_{q}. \biggl [ \begin{matrix} k \\ r \end{matrix} \biggl ]_{q}
\end{gather}
From the total of the Counts, $\texttt{Count}=\texttt{Count}_{U}+\texttt{Count}_{V}+\texttt{Count}_{*}$,  the bounds can be computed. It follows that $U=\texttt{span} \{g_{i},x_{U_{i}} \}$, $V=span\{g_{i},y_{V_{i}} \}$ and $U \cap V= \texttt{span} \{g_{i} \}$
\end{proof}
\begin{remark}
 The syndrome decoding problem becomes \\ $H^{'}x^{T}=\sum_{l=1}^{n}\sum_{j=1}^{k}\alpha_{ij}H_{l}^{'}V_{j}=0$, we now have \\
\begin{equation}
  \text{Prob}(U \cap V)= \frac{q^{\frac{k-1}{2}}\biggl ( \begin{matrix}n \\ k \end{matrix} \biggl )_{q}}{q^{\frac{k(k-r)}{2}}\biggl ( \begin{matrix}k \\ r \end{matrix} \biggl )_{q}} \propto q^{\frac{k(k-r)}{2}(n-k)} 
\end{equation}
This results in a complexity of $O(\frac{(n-k)^{2}}{2}q^{\frac{k(k-r)}{2}(n-k)}) $.
\end{remark}
\begin{theorem}
if the dimension of the vector space $\forall d \leq 2$, then the complexity of basis enumeration is given by $ \sum_{\alpha=1}^{d}\begin{pmatrix}n \\ l \end{pmatrix}\begin{pmatrix}\alpha \\ n \end{pmatrix}^{d}\biggl ( 1- \frac{\alpha}{n} \biggl )^{n-l}x^{d}$ .
\end{theorem}
Theorem 4 gives a closed form expression for the average number of iterations

\section{Failure probability and Complexity Analysis} We present numerical results on the optimization of plucker set decoding to the totally non negative grassmannian. In order to compare the results with code in the Hamming metric, we optimized our implementation to use information sets rather than plucker cooridnates. It is also important we feed the algorithm with as much sets as possible to make the iteration process smooth and efficient. At this juncture it is important to reiterate that simulations of these kind has huge impact on the memory resources of the computing device deployed. In these experiments we used AMD Ryzen 3 2200U laptop with Radeon Vega Mobile Gfx graphic card with processor speed of 2500MHz, 2 cores, 4 logical processors and clock speed of 2.5GHz. Due to the limitation of the memory, the experiments were conducted with little amount of code sizes. However, these experiments can be scaled up without much impact on the result analysis.
\subsection{Probability of failure}In this section, the results of experiments on the probability of decryption failure while using the solution to the syndrome decoding problem to recover the information sequence from totally non negative Grassmannian is presented and compared with the probability of solving the problem using an LDPC code in the Hamming metric. This process was carried out by optimizing the implementation \cite{guo2019some} for this purpose. Theoretical analysis on the comparison between two codes has been studied(ref). We go further than this by experimentally analysing the implication of this comparison on the security of a code based cryptosystem. We can recall the importance of this property on the semantic security of Indistinguishability for a Chosen ciphertext attack. This is because of the negligible error patterns present in each vector space. The lower this probability, the higher chance of the quantum adversary to distinguish between random instances of the ciphertext. In this experiment, we set the number of information sets $2^{l}$, for each level of security under investigation were $l$ is the number of indices of the information set. For 128-bit security level, we set $T=32,768$ and the result is shown in Fig. 3, for 256-bit we set the number of information sets as $l=1048576$ and the result is shown in Fig 4, for 512-security level we set the number of information sets as $T=33,554,432$ and the result is shown in Fig.5,finally for 1024-security level, we set the number of information sets to $T=1073741824$ and the result is shown in Fig. 6. The standard deviation of the distribution $\sigma$ for all security levels is varied from $0.30$ to $0.85$ for cryptography purposes. To compute the amount of Gaussian elimination operation carried out, we use the formula $\frac{1}{2}(n-k)k^{2}$, this is shown in Table 1. This formula relates the number of information sets $T$ to the Gaussian decomposition operations. It can be see from Table 1. that the Gaussian decomposition increases as the security level increases. This is due to size of the information set for each security level which is bounded by $\geq n-k$. The reason for this is to limit the frequent failure of the algorithm due to its probabilistic approach at examining the codewords. However, this comes at a great computational cost. Furthermore, It can be seen that the failure probability of the Non-negative Grassmannian code is smaller than the failure probability of the LDPC code. The implication of this is that the Non-negative Grassmannian code based cryptosystem is more secured than the LDPC code based cryptosystem under the IND-CCA model. This is because in the IND-CCA model, the probability error must be negligible in order for the probability polynomial adversary to find it hard to be able to distinguish a secret sampled from a theoretical distribution from that sampled from an arbitrary distribution. In Fig1. at a standard deviation of $0.50$, the failure probability of the Non-negative Grassmann code is less than that of the LDPC code by 1.18 percent, In Fig2. at a standard deviation of $0.50$, the failure probability of the Non-negative Grassmann code is less than that of the LDPC code by 3.23 percent. In Fig3. at a standard deviation of $0.50$, the failure probability of the Non-negative Grassmann code is less that of the LDPC code by 2.34 percent and finally in Fig.4 at a standard deviation of $0.50$, the failure probability of the Non-negative Grassmann code is less than that of the LDPC code by 3.17 percent.
As the security level increases, the size of the intersection array increases which induces some level of randomness on the plucker coordinates and in the process expanding the probability that a zero error pattern is contained in an arbitrary information subspace. This can be seen in the reduction in the error floor as the security level increases.

\begin{table*}[htp!]
 \centering
   \caption{Gaussian decomposition Operations as a function of Security level}
       \begin{tabular}{c c}
       \hline
         \multicolumn{1}{c}{Security level} & \multicolumn{1}{c}{Gaussian Decomposition}  \\ \cline{1-2}
          128  & 131072   \\
          256 & 1048576   \\
          512 & 8388608 \\
         1024 & 67108864 \\

\end{tabular}
\end{table*}   
    
\begin{table*}[htp!]
    \centering
     \caption{Comparison with Parameters in the Rank metric}
    \begin{tabular}{c c c c c c c c}
    \hline
        \multicolumn{1}{c}{\textbf{n}} & \multicolumn{1}{c}{\textbf{k}} &\multicolumn{1}{c}{\textbf{m}} &\multicolumn{1}{c}{\textbf{q}} 
        & \multicolumn{1}{c}{\textbf{w}} & \multicolumn{1}{c}{\textbf{Security}}& \multicolumn{1}{c}{\textbf{}} \\
         \hline
        67  & 7 & 89 & 2 & 5 & 128 &  \cite{melchor2011new} \\
        100 & 80 & 96 & 2 & 5 & 192 & \cite{gaborit2017identity} \\
        100 & 80 & 96 & 2 & 5 & 192 & \cite{chang2018revocable} \\
        67 & 22 & 71 & 2 & 11 & 133 & \cite{lau2018new} \\
        110 & 7 & 18 & 2 & 12 & 128 & This work \\
    \end{tabular}
    \label{tab:my_label}
\end{table*}

\begin{figure}[!htbp]
    \centering
    \includegraphics[width=9cm, height=7cm]{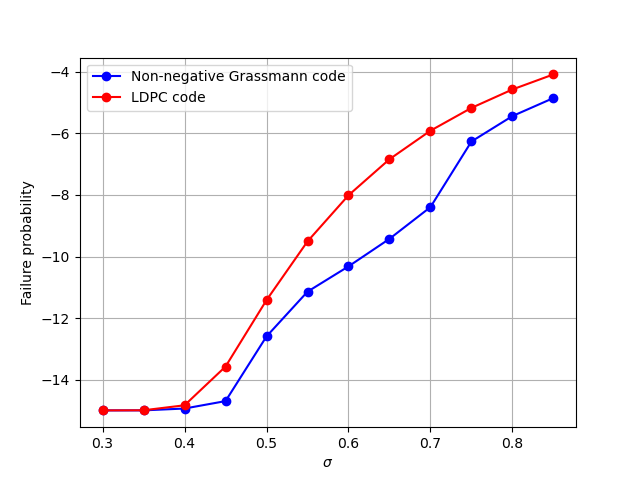}
    \caption{Probability of failure for 128-bit security, security parametr $l=15$}
    \label{fig:my_label}
\end{figure}
\begin{figure}[!htbp]
    \centering
    \includegraphics[width=9cm, height=7cm]{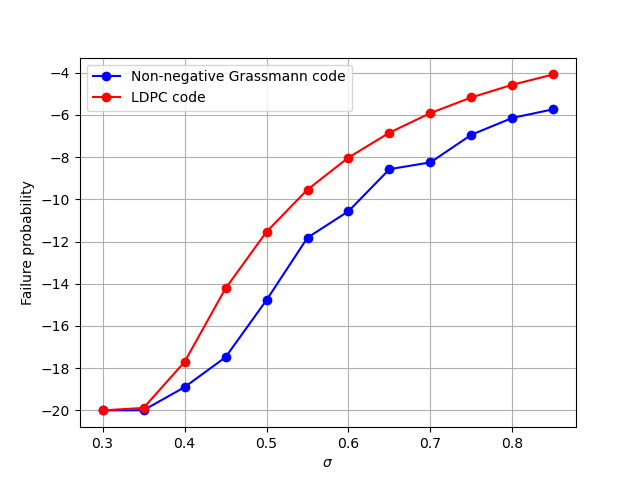}
    \caption{Probability of failure for 256-bit security,security parameter $l=20$}
    \label{fig:my_label}
\end{figure}
\begin{figure}[!htbp]
    \centering
    \includegraphics[width=9cm, height=7cm]{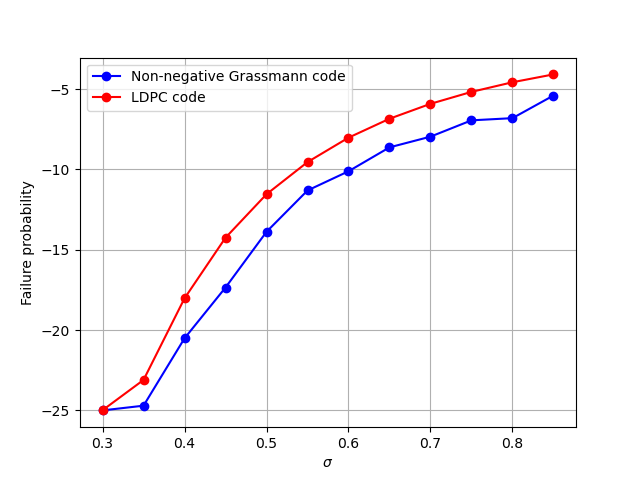}
    \caption{Probability of failure for 256-bit security,security parameter$l=25$}
    \label{fig:my_label}
\end{figure}
\begin{figure}[!htbp]
    \centering
    \includegraphics[width=9cm, height=7cm]{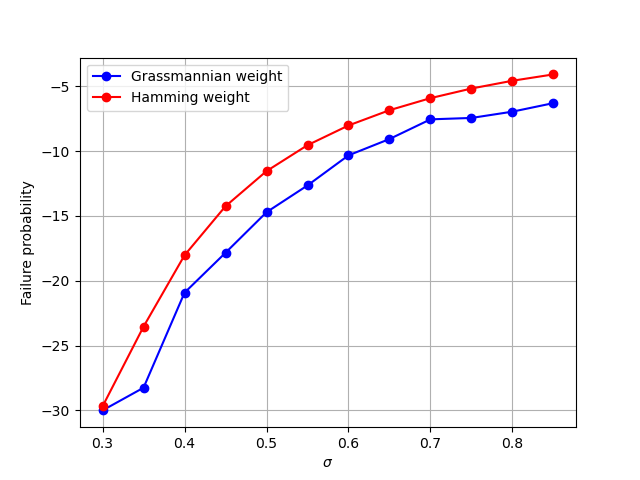}
    \caption{Probability of failure for 1024-bit security, security parameter $l=30$}
    \label{fig:my_label}
\end{figure}

\subsection{Complexity}
In this section we optimized the implementation \cite{beullens2020not} to test the cost of iterating over the rows of the Non negative Grassmann code in the Grassmann metric as compared to the LDPC code in the Hamming metric with increase in code length. The results are presented in Fig. 6 for finite field of characteristic $2$ and in Fig. 7 for a finite field of characteristic $2$ and extension $2$. From the result it can be seen that cost of iterating over rows of the Non negative Grassmann code is higher than of the LDPC code with increasing code length. At a code length of $n=100$, the complexity of row operations is higher by 5.81 percent. This shows that Non negative Grassmann code based cryptosystem is stronger against ISD attack than LDPC code. This is good for quantum security. In Fig. 7, the field size was extended by $2$ and a difference of 29.4 percent was recorded. The huge difference is a result of the large size of the coefficients of the polynomial linear equations with variable $q$, the field size which in turn increases the size of the basis of $k+1$ subspaces of dimension $n=1$.
\begin{figure}[!htbp]
    \centering
    \includegraphics[width=9cm, height=7cm]{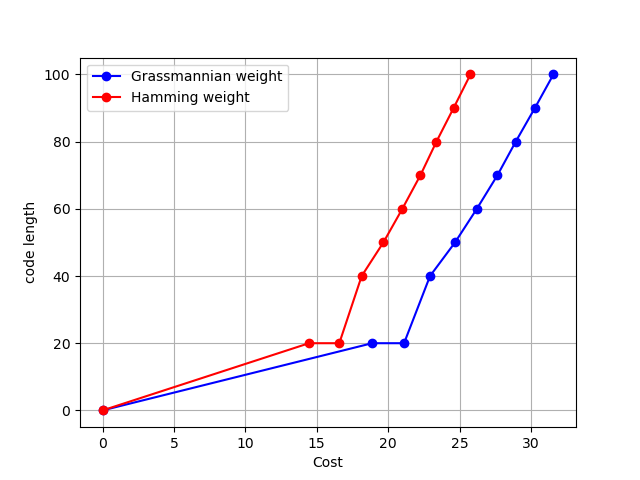}
    \caption{Cost of row ISD operations,field size $q=2$}
    \label{fig:my_label}
\end{figure}

\begin{figure}[!htbp]
    \centering
    \includegraphics[width=9cm, height=7cm]{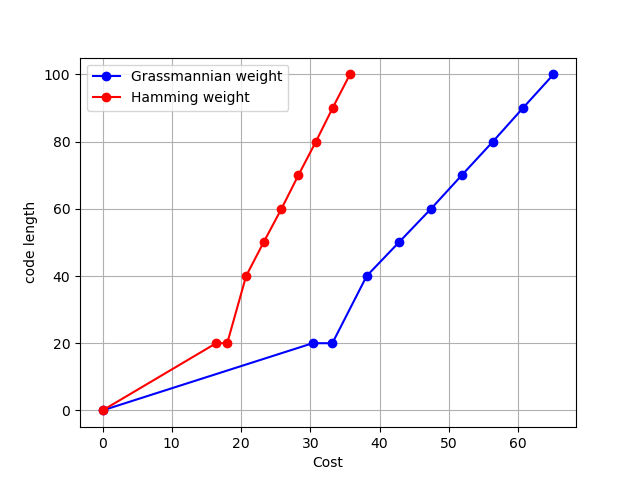}
    \caption{Cost of ISD row operations,field size $q=2^{2}$}
    \label{fig:my_label}
\end{figure}
Quantum security is obtained by dividing the security bits by 2, that means for 128 bit security the equivalent quantum security is 56 bits and to make the density of the decodable syndrome close to 1, parameters must satisfy \cite{bernstein2009introduction}. The Grassmannian weight of the Trapdoor function should be large enough to make cryptanalysis through a structural process difficult. The data size and computational time are linear in $log q$ while the complexity of combinatorics are polynomial on $q$ making it difficult to break encryption key. The decoding error with failure probability is equivalent $\frac{1}{q^{l^{'}-2wr+1}}$ \cite{gaborit2017identity} and the key size increase inversely to an increase in the probability of the decoding failure. In the presence of cyclic vectors, classical attacks makes it possible to obtain the plucker coordinates of the permuted codewords. In Table 2 we give suggested parameters were $n$ is the code length, $k$ is the code dimension, $m$ is the degree of extension field, $q$ is the prime, $w$ is the error weight  which is compared to other parameters from related works. The works compared in the table were variants of ISD employed in cryptanalyzing Code based crytpography in the rank metric. From the complexity derived from Theorem 3 and Remark 2, it can be deduced that the complexity of the ISD decomposition on the input of the proposed parameters is $2^{23}$ which is below the claimed security level of $2^{128}$. This shows that the complexity of our approach depends on the size of the plukcer coordinates as derived from proposition 2.

\section{Conclusion}The syndrome decoding problem as a computationally hard primitive has been used in code based cryptosystem to secure information systems from quantum based solutions.In this paper, we generalize solution to the problem using Information set decoding to the Grassmann metric for codes associated with the totally non negative Grassmanninan. A new theory linking the planar structure of the totally non negative Grassmannian to Tanner graph like construction was developed using the concept of boundary measurement map was developed. The bounds on the parameters such as the size of the information subspace and intersection array  of the new constructed Non-negative grassmann codes was derived. 
Thereafter a variant of Information set decoding based on decomposing the Generator matrix into positroid cells using Gaussian elimination to find linearly dependent subsets of the plukcer coordinates with minimal non-zero coordinates  and in which the  the maximal minor is totally positive was presented. Finally numerical results presented showed that the Non negative Grassmann code had a low decoding probability of failure when compared with an LDPC code. This implies that the error floor of the LDPC code is higher than that of the Non-negative Grassmann code. Also, for increase in the code length, the decoding cost for the totally non negative Grassmann code was higher than the LDPC code. This validates the theory of the Non negative Grassmann code in the Grassmann metric more Indistinguishable secure under the Chosen ciphertext model when compared to the LDPC code in the Hamming metric. Due to its robust security credentials, we recommend this code to construct future post quantum encryption schemes.

\printbibliography

@article{prange1962use,
  title={The use of information sets in decoding cyclic codes},
  author={Prange, Eugene},
  journal={IRE Transactions on Information Theory},
  volume={8},
  number={5},
  pages={5--9},
  year={1962},
  publisher={IEEE}
}

@inproceedings{stern1988method,
  title={A method for finding codewords of small weight},
  author={Stern, Jacques},
  booktitle={International Colloquium on Coding Theory and Applications},
  pages={106--113},
  year={1988},
  organization={Springer}
}

@article{etzion2012codes,
  title={Codes and designs related to lifted MRD codes},
  author={Etzion, Tuvi and Silberstein, Natalia},
  journal={IEEE Transactions on Information Theory},
  volume={59},
  number={2},
  pages={1004--1017},
  year={2012},
  publisher={IEEE}
}

@article{mceliece1978public,
  title={A public-key cryptosystem based on algebraic},
  author={McEliece, Robert J},
  journal={Coding Thv},
  volume={4244},
  pages={114--116},
  year={1978}
}

@inproceedings{chang2018revocable,
  title={Revocable identity-based encryption from codes with rank metric},
  author={Chang, Donghoon and Chauhan, Amit Kumar and Kumar, Sandeep and Sanadhya, Somitra Kumar},
  booktitle={Cryptographers’ Track at the RSA Conference},
  pages={435--451},
  year={2018},
  organization={Springer}
}

@inproceedings{chabaud1996cryptographic,
  title={The cryptographic security of the syndrome decoding problem for rank distance codes},
  author={Chabaud, Florent and Stern, Jacques},
  booktitle={International Conference on the Theory and Application of Cryptology and Information Security},
  pages={368--381},
  year={1996},
  organization={Springer}
}

@inproceedings{gaborit2014ranksign,
  title={RankSign: an efficient signature algorithm based on the rank metric},
  author={Gaborit, Philippe and Ruatta, Olivier and Schrek, Julien and Z{\'e}mor, Gilles},
  booktitle={International Workshop on Post-Quantum Cryptography},
  pages={88--107},
  year={2014},
  organization={Springer}
}

@article{gaborit2015complexity,
  title={On the complexity of the rank syndrome decoding problem},
  author={Gaborit, Philippe and Ruatta, Olivier and Schrek, Julien},
  journal={IEEE Transactions on Information Theory},
  volume={62},
  number={2},
  pages={1006--1019},
  year={2015},
  publisher={IEEE}
}

@inproceedings{kachigar2017quantum,
  title={Quantum information set decoding algorithms},
  author={Kachigar, Ghazal and Tillich, Jean-Pierre},
  booktitle={International Workshop on Post-Quantum Cryptography},
  pages={69--89},
  year={2017},
  organization={Springer}
}

@incollection{bernstein2009introduction,
  title={Introduction to post-quantum cryptography},
  author={Bernstein, Daniel J},
  booktitle={Post-quantum cryptography},
  pages={1--14},
  year={2009},
  publisher={Springer}
}

@article{melchor2011new,
  title={A new efficient threshold ring signature scheme based on coding theory},
  author={Melchor, Carlos Aguilar and Cayrel, Pierre-Louis and Gaborit, Philippe and Laguillaumie, Fabien},
  journal={IEEE Transactions on Information Theory},
  volume={57},
  number={7},
  pages={4833--4842},
  year={2011},
  publisher={IEEE}
}

@inproceedings{gaborit2017identity,
  title={Identity-based encryption from codes with rank metric},
  author={Gaborit, Philippe and Hauteville, Adrien and Phan, Duong Hieu and Tillich, Jean-Pierre},
  booktitle={Annual International Cryptology Conference},
  pages={194--224},
  year={2017},
  organization={Springer}
}

@article{lau2018new,
  title={A new technique in rank metric code-based encryption},
  author={Lau, Terry and Tan, Chik},
  journal={Cryptography},
  volume={2},
  number={4},
  pages={32},
  year={2018},
  publisher={Multidisciplinary Digital Publishing Institute}
}

@article{franco2014bipartite,
  title={Bipartite field theories, cluster algebras and the Grassmannian},
  author={Franco, Sebastian and Galloni, Daniele and Mariotti, Alberto},
  journal={Journal of Physics A: Mathematical and Theoretical},
  volume={47},
  number={47},
  pages={474004},
  year={2014},
  publisher={IOP Publishing}
}

@article{postnikov2009matching,
  title={Matching polytopes, toric geometry, and the totally non-negative Grassmannian},
  author={Postnikov, Alexander and Speyer, David and Williams, Lauren},
  journal={Journal of Algebraic Combinatorics},
  volume={30},
  number={2},
  pages={173--191},
  year={2009},
  publisher={Springer}
}

@article{koetter2008coding,
  title={Coding for errors and erasures in random network coding},
  author={Koetter, Ralf and Kschischang, Frank R},
  journal={IEEE Transactions on Information theory},
  volume={54},
  number={8},
  pages={3579--3591},
  year={2008},
  publisher={IEEE}
}

@article{berlekamp1978inherent,
  title={On the inherent intractability of certain coding problems (corresp.)},
  author={Berlekamp, Elwyn and McEliece, Robert and Van Tilborg, Henk},
  journal={IEEE Transactions on Information Theory},
  volume={24},
  number={3},
  pages={384--386},
  year={1978},
  publisher={IEEE}
}

@book{andrews1986q,
  title={$ q $-Series: Their Development and Application in Analysis, Number Theory, Combinatorics, Physics and Computer Algebra: Their Development and Application in Analysis, Number Theory, Combinatorics, Physics, and Computer Algebra},
  author={Andrews, George E},
  number={66},
  year={1986},
  publisher={American Mathematical Soc.}
}

@article{tanner1981recursive,
  title={A recursive approach to low complexity codes},
  author={Tanner, R},
  journal={IEEE Transactions on information theory},
  volume={27},
  number={5},
  pages={533--547},
  year={1981},
  publisher={IEEE}
}

@inproceedings{gabidulin2015subspace,
  title={Subspace Network Codes with Large Cardinality},
  author={Gabidulin, EM and Pilipchuk, NI},
  booktitle={2015 International Conference on Engineering and Telecommunication (EnT)},
  pages={10--13},
  year={2015},
  organization={IEEE}
}

@article{ryan1990minimum,
  title={The minimum weight of the Grassmann codes C (k, n)},
  author={Ryan, Charles T and Ryan, Kevin M},
  journal={Discrete applied mathematics},
  volume={28},
  number={2},
  pages={149--156},
  year={1990},
  publisher={Elsevier}
}

@inproceedings{varshamov1957evaluation,
  title={The evaluation of signals in codes with correction of errors},
  author={Varshamov, Rom Rubenovich},
  booktitle={Doklady Akademii Nauk},
  volume={117},
  number={5},
  pages={739--741},
  year={1957},
  organization={Russian Academy of Sciences}
}

@article{guo2019some,
  title={Some cryptanalytic and coding-theoretic applications of a soft stern algorithm},
  author={Guo, Qian and Johansson, Thomas and M{\aa}rtensson, Erik and Wagner, Paul Stankovski},
  journal={Advances in Mathematics of Communications},
  volume={13},
  number={4},
  pages={559},
  year={2019},
  publisher={American Institute of Mathematical Sciences}
}

@techreport{beullens2020not,
  title={Not enough LESS: An improved algorithm for solving Code Equivalence Problems over Fq},
  author={Beullens, Ward}
}

@article{etzion2019grassmannian,
  title={Grassmannian codes with new distance measures for network coding},
  author={Etzion, Tuvi and Zhang, Hui},
  journal={IEEE Transactions on Information Theory},
  volume={65},
  number={7},
  pages={4131--4142},
  year={2019},
  publisher={IEEE}
}

\newpage

\end{document}